\documentclass[iop]{emulateapj}
\usepackage{amsmath}
\usepackage{graphicx}
\usepackage{rotating}
\usepackage{xcolor}
\newif\ifAMStwofonts
\AMStwofontstrue
\usepackage{ulem}  
\usepackage{mathtools} 

\def\arcdeg{\hbox{$^\circ$}}

\def\ga{\mathrel{\hbox{\rlap{\hbox{\lower4pt\hbox{$\sim$}}}\hbox{$>$}}}}
\def\la{\mathrel{\hbox{\rlap{\hbox{\lower4pt\hbox{$\sim$}}}\hbox{$<$}}}}

\shorttitle{Timing Relativistic Binary Pulsar PSR B1913+16}

\shortauthors{J. M. Weisberg  and Y. Huang}

\begin{document}

\title{ Relativistic Measurements from Timing the Binary Pulsar \\
  PSR B1913+16}

\author{J. M. Weisberg and Y. Huang}
\affil{Department of Physics and Astronomy, Carleton College, Northfield, MN 55057}
\email{jweisber@carleton.edu, orcid  https://orcid.org/0000-0001-9096-6543}

\begin{abstract}
We present relativistic analyses of 9257 measurements of times-of-arrival from the first
binary pulsar, PSR B1913+16, acquired over the last thirty-five years. The determination
of the ``Keplerian'' orbital elements plus two relativistic terms completely characterizes
the binary system, aside from an unknown rotation about the line of sight; leading to a
determination of the masses of the pulsar and its companion: 
$1.438  \pm  0.001\  {\rm M}_{\sun}$ and $1.390  \pm 0.001\   {\rm M}_{\sun}$, respectively.  
In addition, the complete system characterization allows the creation of tests of relativistic
gravitation by comparing measured and predicted sizes of various relativistic phenomena.
We find that the ratio of observed orbital period decrease due to gravitational wave damping 
(corrected by a kinematic term) to  the general relativistic prediction, is    
$0.9983 \pm 0.0016$;
thereby confirming the existence and strength of gravitational
radiation as predicted by general relativity. For the first time in this system, we 
have also successfully measured the two parameters characterizing  the Shapiro 
gravitational propagation delay, and find that their values are consistent with 
general relativistic predictions.  We have also measured for the first time in any system
the relativistic shape correction to  the   elliptical orbit, $ \delta_{\theta}$, although its 
intrinsic value is obscured by  currently unquantified pulsar emission beam aberration.  
We have also marginally measured the time derivative of the projected semimajor axis,
which, when improved in combination with  beam aberration
 modelling from 
geodetic precession observations, should ultimately constrain the pulsar's moment of inertia.
\end{abstract}

\slugcomment{Accepted by  APJ 2016 Jun 1}


\setcounter{footnote}{0}

\section{Introduction}

Pulsar B1913+16 was the first binary pulsar discovered \citep{ht75}. The system consists 
of two neutron stars 
(one an observed pulsar) orbiting in a very tight, highly eccentric orbit,   and it remains 
one of the best for studying relativistic gravitation 
(Weisberg \& Taylor 1981; Taylor \& Weisberg 1982, 1989; and Weisberg, Nice, \& 
Taylor 2010, hereafter WNT).
 In this paper, we update  WNT with the 
addition of post-2006 data and with further  relativistic timing analyses. 
The addition of significant quantities of data acquired with modern data-acquisition
devices has enabled us to measure several additional relativistic phenomena for the
first time in this system, while also refining previously-measured ones.  Among the
 parameters newly measured with various degrees of accuracy are the Shapiro gravitational 
 propagation delay, a relativistic correction to the elliptical orbital shape, and the time derivative of
 projected pulsar semimajor axis.   All
of the the data used in this study are published with this paper, while our analysis software
is published in an online repository. 

We describe the observations used  in this work in \S\ref{sec:data}, while
\S\ref{sec:relanaly} delineates the scope and methods of our relativistic analyses of these data.  
The results of our  fits to the data are explained in \S\ref{sec:results}, and their applications
for tests of relativistic gravitation are discussed in \S\ref{sec:tests}. We conclude in \S\ref{sec:concl} 
by summarizing   our work  and placing it in the context of results from other relativistic 
binary pulsar systems. 

\section{Data}
\label{sec:data}

The data for our analyses consists of  9257 pulse times of arrival (TOAs) derived from
five-minute integrations of the pulsar signal  at
frequencies near 1400 MHz measured at Arecibo Observatory from 1981 - 2012.  
The parameters
of the various observing systems and the number of TOAs from each through epoch 2006 are
tabulated in WNT; WAPP spectrometer observations since then have  added another 1652
TOAs to the total, each acquired by  three WAPPs deployed 
simultaneously at approximately contiguous 100-MHz bands near 1400 MHz.

Geodetic precession of the pulsar spin axis has induced pulse profile changes  
\citep{wet89,k98,wt02,cw08} that have lately grown increasingly larger, presumably as
our line of sight approaches the edge of the pulsar beam.  Nevertheless, for purposes of 
uniformity, we use only a single  profile template while finding TOAs for all WAPP data. 
This  procedure induces  time offsets  into our TOA  dataset 
between different sessions and frequencies, 
which have grown to a level where they should be compensated for.
To do so, we adopted the following process. 
First, we formed a pulse profile at each frequency band for each two-week session.  
Each resulting ``session - band'' standard
profile has much greater  $S/N$ than 
does a single five - minute integration, while still being short enough to avoid the secular
changes we are trying to measure.
Next, we measured the offset of the midpoint of this session - band standard profile 
with respect to the  grand standard profile. (The midpoint
is assumed to correspond to a fixed longitude on the pulsar regardless of profile shifts).

Then we fitted out	
a ``primary'' linear model of the profile offsets 
as a function of time at each band.   In this fashion,
we provided an empirically determined, model independent, first-order TOA correction
that accounts for the effects of profile changes, thereby minimizing
  the long-term effects of profile shifts 
that might be mistaken as the  signature of other phenomena. We next   fitted  the timing 
model to all such  ``primarily-offsetted'' TOAs and chose 
the resulting dispersion measure as our nominal value.  
Finally, secondary offsets were determined through single session fits, where the 
dispersion measure was fixed at its nominal value and the 
residual offset of each band was fitted for and then removed. 
This process ensures that the infinite-frequency TOAs calculated from each band  
and session are self-consistently de-dispersed\footnote{This procedure also 
absorbs TOA variations 
induced by DM fluctuations at the  levels and timescales expected from studies
of millisecond pulsars \citep{yet07}. } and offsetted.

To verify that the above profile-variation-correction
process does not contaminate our parameter measurements, we also employed  
 an alternate approach, 
fitting for an offset for each band in each session simultaneously with all other
parameters   \citep{det13}. This procedure yields parameters that agree with our 
method to within  1$\sigma$ for all parameters, 
suggesting that our measurements are robust with respect to the methods 
used to remove profile-shift-induced timing offsets. 

\section{Relativistic Analysis of TOAs}
\label{sec:relanaly}

Using  an augmented version  of the TEMPO software program, 
we fitted the relativistic timing model  of \citet[hereafter DD]{dd86}; or, in certain cases,  
the DD model augmented by the \citet[hereafter FW]{fw10} Shapiro parametrization 
(see \S\ref{sec:fitting}),\footnote{See \url{http://sourceforge.net/projects/TEMPO/}  for our 
augmented  version of TEMPO, which contains our fitting routine   for the FW
Shapiro parameters in high-eccentricity binaries.}  to our TOAs. 
 In these models, the pulsar signal encounters several  distinct types of delays on 
its journey from the orbiting pulsar to the solar system barycenter, such that the 
infinite-frequency 
pulse arrival  time at the solar system barycenter, $t_{\rm ssbc}$, is given by:
\begin{equation}
\begin{multlined}
t_{\rm ssbc}= D^{-1}[
T+\Delta_{\rm Roemer}(T)+\Delta_{\rm Einstein}(T) +\\ \\
\Delta_{\rm Shapiro}(T)+\Delta_{\rm Aberration}(T)   ] ;  
\end{multlined}
\label{eqn:deltas}
\end{equation}
where  each delay is a function of the pulsar proper time of pulse emission, $T$, and whose 
details depend on a 
number of physical parameters that are fitted for. (The Doppler factor $D$ accounts for the
relative motion of the solar system and binary system barycenters.)  The various
terms in Eq. \ref{eqn:deltas} are detailed in  
DD and \citet[hereafter DT92]{dt92}, and we will comment
further  on the last two terms of Eq.  \ref{eqn:deltas} in the following two sections.

Among the  fitted parameters, we determined
improved values of the pulsar spin and orbital parameters that were published in WNT,
plus a number of   new ones.  For the first time, we have successfully fitted 
for the Shapiro (1964) gravitational propagation delay while also placing constraints 
on two additional ones: 
a relativistic correction to the  quasi-elliptical shape of the orbit and the shrinkage 
rate of its projected  semimajor axis,  as described further below.

The improved, previously fitted parameters include the pulsar spin frequency  
and derivative(s) $f$,  $\dot f$, \dots; five
``Keplerian'' orbital elements defined as projected pulsar semimajor axis 
$x \equiv a_{1} \sin i$  where $i$ is the orbital inclination, 
orbital period $P_{\rm b}$, eccentricity
$e$, reference epoch $T_0$, the reference epoch's longitude of periastron $\omega_0$; 
relativistic ``post-Keplerian'' parameters defined  as mean rate of 
periastron advance $\langle \dot \omega \rangle$,  gravitational redshift - time dilation term 
$\gamma$, and orbital period derivative $\dot P_{\rm b}$.   

The  newly fitted post-Keplerian parameters include the following:   (i) two Shapiro delay  terms    
 called    shape $s$ and range $r$ in the  
 DD parametrization, or two (different) quantities   $\varsigma$ and $h_3$  in  
 the alternate FW  parametrization;   (ii) the orbital 
elliptical shape correction parameter $\delta_{\theta}^{\mbox{obs}}$  (to our 
knowledge never previously fitted for in any binary system), which appears in DT92's full
expression  for Eq. \ref{eqn:deltas}'s
Roemer term\footnote{Before it can be utilized for tests of relativity,
the $\delta_{\theta}^{\mbox{obs}}$ parameter
must be corrected for a comparable aberration term which  is currently
 undeterminable (see \S \ref{sec:shapeobs}). }; 
 and (iii) $\dot{x}^{\rm obs}$ and $\dot{e}^{\rm obs}$, the observed  time derivatives of $x$ 
and $e$. All of the new parameters
are  discussed in greater detail in \S\ref{sec:fitting}-\S\ref{sec:otherparams}; while 
the fit results 
for  both  old and new parameters are described in \S\ref{sec:results}, and 	
relativistic tests resulting from these measurements are discussed in \S\ref{sec:tests}.
A set of TEMPO input files containing  input parameters and the TOAs is available 
as an online companion to this paper, on arXiv, and at zenodo:  \dataset[10.5281/zenodo.54764]{http://dx.doi.org/10.5281/zenodo.54764}.

\subsection{Fitting for the Shapiro Gravitational Propagation Delay 
via two Different Parametrizations}
\label{sec:fitting}

Until this work, we have  been unable to explicitly  measure the two Shapiro 
gravitational propagation 
delay terms  that characterize Eq. \ref{eqn:deltas}'s
$\Delta_{\rm Shapiro}(T)$, owing 
to their relatively small timing signature and to their covariance.    The two
terms are identified as  $s$ and $r$  in DD;
while  FW recently developed an alternate parametrization of the phenomenon, 
wherein their
two fitted parameters,  $\varsigma$ and $h_3$, are orthonormal.  Our 
software implementation  of their parametrization for high-eccentricity pulsars is available in
our augmented version of TEMPO.$^2$  

The measurements of either pair of Shapiro parameters,  $(s,r)$ or $(\varsigma,h_3)$, can be
utilized in either of two different manners, as described below.

First, if general relativity is assumed to be the correct 
theory of gravitation, then either pair of  Shapiro measurables  can 
be utilized as  independent constraints on the orbital inclination and binary companion mass.
We summarize the theory here, and then apply it in \S\ref{sec:shapiromeas}.

 In the  DD formulation, $s$ and $r$ translate directly 
 into  $\sin i$ and $m_{2}$ (the companion mass), respectively:
\begin{equation}
\sin i = s,
\label{eqn:siniequalss}
 \end{equation}
and
 \begin{equation}
 m_{2} =  \left( \frac{c^3}{ \rm G} \right) \ r =
 \left(  \frac{r}{\ \rm T_\sun}  \right) \rm M_\sun,
\label{eqn:m2equalsr}
 \end{equation} 
 with $c$ the speed of light, G the Newtonian gravitational constant,  
 T$_\sun= $\ G M$_\sun/c^3 = 4.925\ 490\ 947 \ \mu s$.
  
The alternate 
FW parametrization of the Shapiro delay gives 
  \begin{equation}
\sin i =\frac{2 \varsigma}{\varsigma^2+1},
 \label{eqn:sifw10}
 \end{equation}
while $m_{2}$ is
a combination of the two measurables $(h_3,\varsigma)$:
 \begin{equation}
 m_{2} =  \left( \frac{c^3}{2 \ \rm G} \right) \ \frac{h_3}{\varsigma^3} =
 \left(  \frac{h_3/\varsigma^3}{\ \rm T_\sun}  \right) \rm M_\sun.
 \label{eqn:mcfw10}
 \end{equation} 

Alternatively, each measured parameter of the Shapiro pair
can be considered to be  an independent test of relativistic gravitation. 
We apply this procedure in  \S\ref{sec:testshap}.

\subsection{Determination of the Relativistic Orbital Shape Correction $\delta_{\theta}$ 
in the  Presence of the Aberration Delay}
\label{sec:deltathetatheory}

In order to  successfully measure the intrinsic value of $\delta_{\theta}$,
which nominally quantifies a relativistic correction to the shape of the 
approximately elliptical orbit in Eq. \ref{eqn:deltas}'s Roemer delay expression, one
must compensate the observed value for the influence of a  phenomenon that 
comparably affects 
TOAs, namely the orbital-phase dependent aberration of the pulsar beam, 
as described by DD and DT92.  Those authors provide a prescription for 
calculating and eliminating
the confounding aberration signature from the observed value of $\delta_{\theta}$, if the 
aberration geometry is known.  In principle, the necessary information can be gleaned from
 studies of profile changes resulting from geodetic precession of the pulsar spin axis 
 \citep{wt02,cw08}.  In this section, we summarize
the theoretical expressions required to quantify $\delta_{\theta}$ and aberration, and 
we will apply
our obervations to these results in \S\ref{sec:shapeobs}.

The  time delay  $\Delta_{\rm Aberration}$ in Eq. \ref{eqn:deltas}, resulting from  
aberration  of the 
rotating pulsar beam, is dependent on the time-variable  transverse 
component of the pulsar's orbital velocity. DD and DT92 
parametrize the instantaneous delay via the aberration parameters $A(t)$ and $B(t)$:
\begin{align}
  \Delta_{\rm Aberration}=&A(t)\{\sin(\omega+A_e(u))+e\sin\omega\}  \nonumber\\
  +&B(t)\{\cos(\omega+A_e(u))+e\cos\omega\},
\end{align}
where $A_e(u)$ is a true-anomaly-like quantity,  and
$A(t)$ and $B(t)$ are dependent on the   precessing spin-axis geometry: 
\begin{equation} 
  A(t) =-\frac{f^{-1}}{P_b}\ \frac{x}{\sin i\ (1-e^2)^{1/2}}\ \frac{\sin \ \eta}{\sin \ \lambda},\\
  \label{eqn:A}
 \end{equation}
\begin{equation} 
  B(t)  =-\frac{f^{-1}}{P_b}\ \frac{x}{\sin i\ (1-e^2)^{1/2}}\ \frac{\cos \ i \cos \eta}{\sin \ \lambda},
\end{equation}
with $\lambda$ and $\eta$  the geodetically-precessing polar angles of the pulsar spin axis
with respect to the line of sight and line of nodes, respectively. (DD and DT92 suggested the
substitution of a single fixed parameter, $A_0$, for the two parameters $A(t)$ and $B(t)$,
because observations at the time suggested
that the spin and orbital angular momenta are aligned.  However, subsequent observations of pulse 
profile changes have shown that this is not the case.)

While the above equations provide a complete  description of the calculation of 
$\Delta_{\rm Aberration}$ at any proper emission time,  DD and DT92 also  provide an alternate
approach  that focuses on aberration parameters that change slowly (on precession timescales) 
as a result of spin-axis 
precession. This procedure, detailed below,  is more closely tailored to   parameters determinable
from TOA analyses. 

DT92 show that aberration will bias  the observed value of the 
relativistic orbital shape parameter $\delta_{\theta}^{\mbox{obs}}$ with respect to its 
intrinsic value $\delta_{\theta}^{\mbox{intr}}$:  
\begin{equation} \delta_{\theta}^{\mbox{obs}}  =  \delta_{\theta}^{\mbox{intr}} - \epsilon_A,
\label{eqn:deltathetaobs}
\end{equation}
where the small parameter $\epsilon_A$ is defined as 
\begin{equation}
\epsilon_A\equiv\frac{A(t)}{x}.
\label{eqn:epsilonA}
\end{equation}

Hence the observational bias can be removed, given  measurements of the aberration parameter $A(t)$
 and the Keplerian quantity $x$. 
The corrected value $\delta_\theta^{\mbox{intr}}$ 
could then serve as an additional test of gravitation theory. 

\subsection{Other Parameters Affected by the Aberration Delay}
\label{sec:otherparams}
In addition to affecting $\delta_\theta$, DT92 show that  aberration also affects the 
observed $x$ and $e$ values.
However, the fractional corrections to $x$ and $e$ are tiny. More interesting is the effect of  geodetic
 spin-axis precession on the  time-derivatives of these parameters, because   DT92 show that they
are potentially measurable.
The precessional motion will cause the aberration geometry to change, resulting
in secular changes to  $A(t)$ and hence to $\epsilon_A$ on precession timescales:
\begin{equation}
\begin{multlined}
\frac{d\epsilon_A}{dt}  = \frac{1}{x} \frac{dA(t)}{dt}= -\frac{f^{-1}}{P_b}\ \frac{1}{\sin i\ (1-e^2)^{1/2}} 
  \ \frac{\Omega_{1}^{\rm geodetic}}  {\sin^2\lambda} \times \\ \\
  (\sin i \cos \lambda \sin 2\eta+\cos i \sin \lambda \cos \eta),
\end{multlined}
\label{eqn:depsilonAdt}
\end{equation}
where $2 \pi/\Omega_{1}^{\rm geodetic}$ is the geodetic precession 
 period of the pulsar spin axis.  The observed, normalized time derivative of $e$ 
results from the sum of two phenomena: 

\begin{equation}
\left(\frac{\dot{e}}{e}\right)^{\mbox{obs}}= \frac{d\epsilon_{A}}{dt} + 
                                \left(\frac{\dot{e}}{e}\right)^{\mbox{GW}},  
\label{eqn:edot}
\end{equation}
where ``GW'' designates effects due to gravitational waves; while
the observed, normalized time derivative of $x$  stems from a  combination of five terms:
\begin{equation}                  
 \begin{multlined}              
  \left(\frac{\dot{x}}{x}\right)^{\mbox{obs}} =   \frac{d\epsilon_{A}}{dt} +
\left(\frac{ \dot{a}_{1} }{{a}_{1}}\right)^{\mbox{GW}} +\\ \\
\left(\cot i \frac{di}{dt}\right)^{\mbox{SO}}
- \mu\cot i \sin (\Theta_{\mu}-\Omega)-\frac{\dot{D}}{D},
\end{multlined}
\label{eqn:xdot}
\end{equation} 
 where ``SO'' refers to spin-orbit coupling.  (See \citet{lk04} for an expression that includes
 additional terms needed for some other binary pulsars.)
 
 The quantity $di/dt$ in the third 
 term of Eq. \ref{eqn:xdot}, 
 resulting from pulsar spin-orbit coupling, is developed here from expressions in DT92:
\begin{align}
\frac{di}{dt}=&\frac{\rm G}{c^2}\left\{  2 + \frac{3} {2}  \frac{m_{2}} {m_{1}}    \right\}
\frac{S_{1} }{a_{\rm R}^3 (1-e^2)^{3/2} } \sin \lambda \cos \eta \nonumber \\
=&\frac{1}{c^2}\left\{  2 + \frac{3} {2}  \frac{m_{2}} {m_{1}}    \right\}
\frac{I_{1}  (2 \pi f)(2 \pi / P_{\rm b})^2}
{(m_{1} + m_{2}) (1-e^2)^{3/2} } \sin \lambda \cos \eta,
\label{eqn:didt}
\end{align}
where  $m_1$ is the pulsar mass,
$a_{\rm R}$ is the semimajor axis of the relative orbit, $S_{1}=I_{1}  (2 \pi f)$  is 
the magnitude of the pulsar spin angular momentum, and $I_{1}$ is its moment of inertia.
The  fourth term of Eq. \ref{eqn:xdot} results from the changing projection of the  line of sight  
onto the 
orbital plane  due to proper motion, with  $\mu$ and 
$\Theta_{\mu}$  respectively the amplitude and  position angle 
of proper motion and  $\Omega$  the position angle of the line of nodes \citep{k96}. The final
term of Eq. \ref{eqn:xdot}, involving changes in the Doppler 
factor  $D$ of Eq. \ref{eqn:deltas}, is caused by the relative 
line-of-sight galactic accelerations of the
solar system and the binary system.

The above equations demonstrate that measurements of $\dot{e}$ or $\dot{x}$, 
along with 
experimental or theoretical determinations of some of the other quantities  appearing therein, 
can usefully constrain yet others.

\section{Results of the Fits}
\label{sec:results}

We fitted the parameters discussed above  to the full set of TOAs, using the TEMPO software, as 
modified by us.$^1$  See Tables  \ref{table:astromfits} and \ref{table:orbfits} for our 
results and their
estimated uncertainties.  The  uncertainties quoted therein represent the standard errors 
from the TEMPO fit (except as noted).  This convention differs from our previous
practice, wherein many uncertainties were instead estimated from fitted parameter variations 
across multiple reasonable fits.  While the old procedure facilitated the incorporation of some 
systematic uncertainties into the error budget; the more stable recent 
instrumental configurations 
appear to minimize such effects.  

Some of the fitted parameters shifted by several $\sigma$ with respect to the values
reported in \citet{wnt10}.    The shifts  can all be attributed to the new incorporation 
of a frequency and time
offset for each WAPP observing session and center frequency in order
to account for geodetic-precession-induced profile changes (see \S\ref{sec:data}), 
and to our new
procedure of fitting for rather than  freezing  at 0 the parameter $\dot{x}$. The latter 
procedure also led
to a significantly larger uncertainty in the fitted value of $\gamma$ and in 
quantities derived therefrom.

\begin{deluxetable}{lll}
\tablecolumns{3}
\tablecaption{Astrometric and Spin Parameters }
\tablehead{
\colhead{Parameter}  &   \colhead{Value\tablenotemark{a}} \\
}
\startdata
$t_0$ (MJD)\tablenotemark{b}\dotfill   & 52984.0 \\  
$\alpha$ (J2000)\dotfill & $19^h 15^m 27\fs 99942(3)$ & \\
$\delta$ (J2000)\dotfill &   16\arcdeg 06\arcmin 27\farcs   3868(5)   & \\ 
$\mu_{\alpha}$ (mas yr$^{-1}$)\dotfill &   $-$1.23(4) &  \\ 
$\mu_{\delta}$ (mas yr$^{-1}$)\dotfill  &   $-$0.83(4) & \\ 
$f$ (s$^{-1}$)\dotfill  & 16.940537785677(3)  & \\
$\dot{f}$ (s$^{-2}$)\dotfill  & $-$2.4733(1) $\times 10^{-15}$ &  \\ 
\cutinhead{Glitch Parameters} 
Glitch epoch (MJD)\dots & 52777(2)       \\
$\Delta f$ (s$^{-1}$)\dotfill & 5.49(3)  $\times 10^{-10}$ \\ 
$\Delta \dot{f}$(s$^{-2}$)\dotfill &  $-$2.7(1) $\times 10^{-18}$ \\ 

\enddata 
\tablenotetext{a}{Figures in parentheses represent formal TEMPO standard errors
in the last quoted digit, except for the glitch parameters. The stated uncertainty in glitch 
epoch results 
from empirically varying the glitch epoch until $\Delta \chi^2$  corresponds to 
the $68 \%$ confidence level; 
the quoted uncertainties in the other glitch parameters were derived from their variations as
the glitch epoch was varied over the chosen range.}
  \tablenotetext{b}{This quantity is the epoch of the next six measurements
  tabulated here.}
 \label{table:astromfits}
\end{deluxetable}

The  astrometric and spin solutions are listed in Table \ref{table:astromfits}. These are quite 
similar to
those given in \citet{wnt10}, except that our longer post-glitch baseline made it clear that the
previously discovered glitch at MJD $\approx 52770$ is better modeled with the addition of a
change in spin frequency $derivative,\  \Delta \dot{f}$. There remains only
one known glitch having a significantly smaller value of $\Delta f / f$ [in globular 
cluster millisecond
PSR B1821-24 \citep{met09}], although several of magnitude 
similar to the one tabulated here are now known. [See the online Jodrell Bank 
Pulsar Glitch Catalogue\footnote{\url{http://www.jb.man.ac.uk/pulsar/glitches.html} }
(Espinoza et al 2011)]. Note that, as with \citet{wnt10}, ten 
higher-order 
spin derivatives were also fitted for to eliminate the effects of timing noise. Their values are 
not shown in the Table as they do not correspond to meaningful physical parameters.  

\begin{deluxetable}{ll}
\tablecolumns{2}
\tablecaption{Orbital Parameters}
\tablehead{
\colhead{Parameter}  &   \colhead{Value\tablenotemark{a}}  \\
}

\startdata

$T_0 $ (MJD)\dotfill & 52144.90097849(3) \\
$x\equiv a_1 \sin i$  (s)\dots & 2.341776(2) \\
$e$ \dotfill &   0.6171340(4)\\
$P_b$ (d)\dotfill  &    0.322997448918(3)\\
$\omega_0$ (deg)\dotfill   &  292.54450(8)\\
$\langle\dot{\omega}\rangle$ (deg / yr)\dotfill & 4.226585(4)\\
$\gamma$ (ms) \dotfill & 0.004307(4)\\
$\dot{P}_b^{\rm obs} $\dotfill  & $-$2.423(1) $\times 10^{-12}$\\
$\delta_{\theta}^{\rm obs}$\dotfill  & 4.0(25) $\times 10^{-6}$ \\
$\dot{x}^{obs}$\dotfill  & $-$0.014(9)  $\times 10^{-12}$ \\
$\dot{e}^{obs} \ $(s$^{-1}$)\dotfill  & 0.0006(7)  $\times 10^{-12}$ \\
\cutinhead{Shapiro Gravitational Propagation Delay Parameters} 
\sidehead{\citet{dd86}  Parametrization} 
$s$ \dotfill  &$ 0.68\substack{+0.10\\-0.06}$\\ 
$r \ (\mu$s)  \dotfill  & $ 9.6\substack{+2.7\\-3.5}$\\
\sidehead{\citet{fw10} Parametrization} 
$\varsigma$  \dotfill  & 0.38(4) \\
$h_3$  \dotfill & 0.6(1) $\times 10^{-6}$  \\
\enddata 
\tablenotetext{a}{Figures in parentheses represent formal TEMPO standard errors
in the last quoted digit.  The DD Shapiro parameters $s$ and $r$, which are
highly covariant, and their uncertainties, were refined through a 
process illustrated in Fig. \ref{fig:shapirochisq}.}
\label{table:orbfits}
\end{deluxetable}

Table \ref{table:orbfits} displays the results of our fit to orbital parameters, including the eight 
final entries, which are fitted  here for the first time in this system. 
Note that the first two of these eight new parameters, namely  
 $\delta_{\theta}^{\rm obs}$ and $\dot{x}^{\rm obs}$ , are   measured
at the marginal $1.5\sigma$ level, while the third, $\dot{e}^{\rm obs}$, is only an upper limit.
All others in this Table, including the new Shapiro terms, are measured with  high confidence.
 In the next sections,
we discuss important orbital measurables, including 
corrections that must be made to 
some of the observed quantities  in order to  determine their intrinsic values.

\subsection{The Observed and Intrinsic Orbital Period Derivative }
\label{sec:obsintPbdot}

The observed orbital period derivative $\dot{P}_{\rm b}^{\rm obs}$, must be corrected by a
term  $\dot{P}_{\rm b}^{\rm  gal}$, resulting from the relative galactic 
accelerations of the solar system and the binary system 
(Damour \& Taylor [1991, hereafter DT91], DT92),
 in order to yield the intrinsic derivative, $\dot{P}^{\rm intr}_{\rm b}$:
\begin{equation}
 \dot{P}^{\rm intr}_{\rm b}=\dot{P}_{\rm b}^{\rm obs} - \dot{P}_{\rm b}^{\rm  gal}.
\label{eq:gal}
\end{equation}
Using galactic parameters of
 $R_0=8.34 \pm 0.16$ kpc and $\Theta_0= 240 \pm 8$ km/s
from \citet{ret14},  a pulsar distance estimate from \citet{wet08},
and the pulsar proper motion from Table \ref{table:astromfits}, we find that 
$\dot{P}_{\rm b}^{\rm  gal}= - (0.025 \pm 0.004)   \times~10^{-12}$. 
Inserting  also $\dot{P}_{\rm b}^{\rm obs}$ from 
Table \ref{table:orbfits} into
Eq. \ref{eq:gal}, we calculate that $ \dot{P}^{\rm intr}_{\rm b}=   
 -(2.398 \pm 0.004) \times 10^{-12} $.
The uncertainty in this result is dominated by the error in $ \dot{P}_{\rm b}^{\rm  gal},$
 which in turn is  set principally by
the pulsar distance  uncertainty.  A VLBA  parallax campaign on the pulsar, currently in 
progress, will hopefully    improve these uncertainties.


\begin{figure}
\vspace{0.7in}
\includegraphics[angle= 90, width=2.8in, trim={0.5in 2.8in 2.6in 0.7in}  ]
{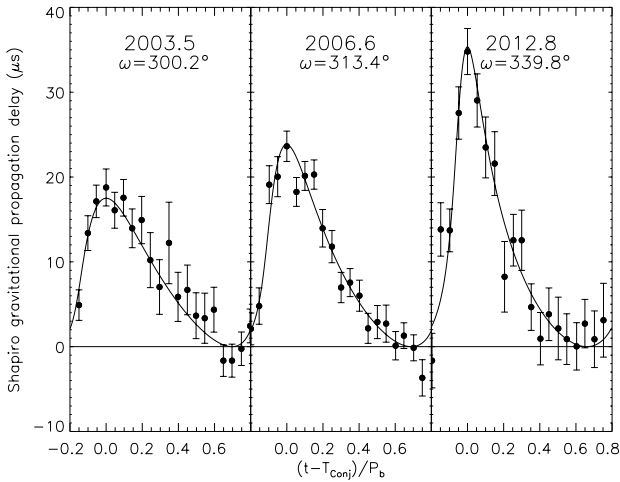} 
\caption{The Shapiro gravitational propagation delay variation around the orbit at three epochs. 
The curve represents the expected delay based on a general relativistic calculation, while the 
points and their error bars result from combining all residuals to a special fit (see text) 
near the given epoch into one of 
20 orbital time bins.  Time is reckoned  with respect to $T_{\rm Conj}$, the
epoch of the pulsar's superior conjunction with the companion.   Each curve peaks at that epoch, 
when the pulsar's  earthbound signals 
plunge most deeply into the companion's gravitational well. The  amplitude and shape of the curves  
evolve, due to relativistic precession of the orbital ellipse, as quantified
by the advancing longitude of periastron $\omega$.}
\label{fig:shapiroovertime}
\end{figure}
\vspace{0.3in}

\subsection{First Successful Measurement of the Shapiro Gravitational Propagation Delay 
Parameters in PSR B1913+16}
\label{sec:shapiromeas}

Because of     relativistic precession of the elliptical orbit, the Shapiro 
delay has recently grown to an amplitude of  $\sim35\ \mu s$ around  the orbit, rendering it 
relatively easily  measurable.  Fig. \ref{fig:shapiroovertime} illustrates the enhancement of the 
Shapiro delay signal  around the orbit over the last dozen years, during which time the WAPP 
receivers also came into use, thereby increasing our observing bandwidth ten-fold.  The
curves in the Figure illustrate a general relativistic calculation of the   expected Shapiro delay 
variation around the orbit,
while the data points are residuals to a TEMPO fit freezing all parameters 
 at their best-fit values, except for the Shapiro parameter $r$ (corresponding in 
General Relativity to the companion mass).    The latter quantity was artificially set to zero
to simulate the absence of the Shapiro delay in the fit.   The  pattern of  
residuals  systematically matches theoretical expectations  for the Shapiro 
delay. 

\begin{figure}
\vspace{0.4in}
\includegraphics[angle= 0,scale=0.43, trim={0.6in 0.4in 0.9in 1.0in,clip}]
{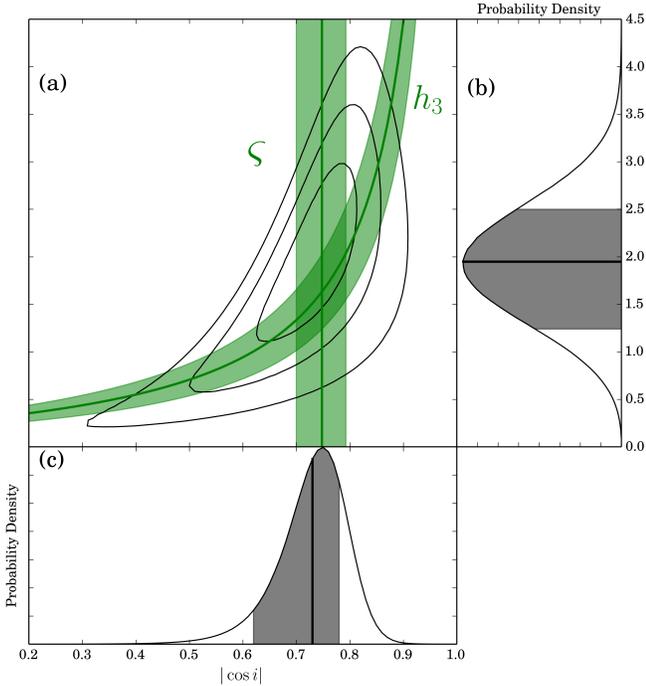} 
\caption{ (a) Measured constraints on  $|\cos i|$ and $m_{2}$ 
resulting  from TEMPO fits for two different parametrizations of the Shapiro gravitational 
propagation delay within the context of  general relativity. The 
\citet{dd86} $s$ and $r$  parameters map directly onto the displayed $|\cos i|
\ [=+\sqrt{1-s^2}\ ]$ 
and $m_{2}\  [ = ( r / \rm T_\sun ) \rm M_\sun ]$ axes, 
respectively;  the black contours   show joint 1, 2, 
and 3$\sigma  \ (\Delta \chi^2  = 2.3, 6.2, 11.8$)  confidence limits on those quantities, 
derived from  a set of 
TEMPO fits  to a large grid of (fixed) ($|\cos i|,m_{2} )$.
The alternate  \citet{fw10} best-fit parameter constraints and their $\pm 1\sigma$ limits  
are shown in green. Their fitted parameter  
$\varsigma$ transforms directly into the displayed $|\cos i|$ axis 
(see Eq. \ref{eqn:sifw10}), 
whereas their $h_3$ parameter  does not map uniquely onto {\it{either}} of the  
axes (see Eq. \ref{eqn:mcfw10}). 
 The marginal distributions in (b) and (c)  result from 
 collapsing the resulting two-dimensional DD 
 probability  distribution  onto the  $|\cos i|$ and
 $m_{2}$ axes respectively, in which the mean (solid black) and the 
1$\sigma$ bounds (grey region) are displayed, yielding 
$|\cos i| = 0.73\  (+0.05, -0.11)$ and
$m_{2}= 1.95 \ (+0.55,-0.71)\  {\rm M}_\sun$ (68.3\% 
confidence).
} 
\label{fig:shapirochisq}
\end{figure}

We have now  successfully determined the two  Shapiro terms in both the 
DD and FW parametrizations
(see Table \ref{table:orbfits}). While the Shapiro delay 
 has been observed in several other binary pulsar systems, these results mark the
first successful detection of a pair of Shapiro measurables in the PSR B1913+16 system.  [The
two DD parameters, $s$ and $r$, had been {\it{jointly}} constrained by \citet{tw89}.
For the ensuing decade, unfavorable orbital geometry rendered its amplitude unmeasurably 
small; while the last decade has seen both improving orbital geometry and advances
in observing instrumentation.]
We  accounted for the significant nonlinear  covariance of the the 
$s$ and $r$ parameters in estimating their values and uncertainties in Table \ref{table:orbfits}, 
using a   process  delineated by \citet{spet02}.
The  procedure is illustrated graphically 
 in Fig.  \ref{fig:shapirochisq}, which also shows the best-fitting FW
Shapiro parameters.  (Tighter constraints on the inclination and
companion mass can be  derived  {\it{indirectly}} from other 
measurements.  See \S\ref{sec:bestmasses}.)

\subsection{Best Determination of Component Masses and Orbital Inclination}
\label{sec:bestmasses}

The measurement of the first seven quantities in Table \ref{table:orbfits} enables the 
precise general relativistic determination of the 
 component masses and orbital inclination. Specifically,
our measurements of $\langle \dot{\omega} \rangle$ and $\gamma$, along with the 
Keplerian elements,
leave only the  two unknowns $m_{1; \langle\dot{\omega}\rangle,\gamma}$ and 
$m_{2;  \langle\dot{\omega}\rangle,\gamma} $ in the following two 
general relativistic equations:

\begin{equation}
\begin{multlined}
\langle\dot{\omega}\rangle = \\
 3 \  {\rm G^{2/3} \  c^{-2} }\
(P_b / 2 \pi)^{-5/3} \  (1-e^2)^{-1} \ (m_{1;  \langle\dot{\omega}\rangle,\gamma} +
 m_{2;  \langle\dot{\omega}\rangle,\gamma} )^{2/3}   \\
= 3  \ {\rm T_\sun^{2/3}  }
(P_b / 2 \pi)^{-5/3} \  (1-e^2)^{-1}  \left(\frac{m_{1;  \langle\dot{\omega}\rangle,\gamma} + 
m_{2;  \langle\dot{\omega}\rangle,\gamma} }  
{\rm M_\sun} \right)^{2/3},
\end{multlined}
\end{equation}
and
\begin{equation}
\begin{multlined}
\gamma  =  {\rm G^{2/3} \  c^{-2} } \ e \ (P_b / 2 \pi)^{1/3} \ m_{2;  \langle\dot{\omega}\rangle,\gamma}  
 (m_{1;  \langle\dot{\omega}\rangle,\gamma} + 
2 m_{2;  \langle\dot{\omega}\rangle,\gamma} )  \times \\  \\
 \ (m_{1;  \langle\dot{\omega}\rangle,\gamma} + m_{2;  \langle\dot{\omega}\rangle,\gamma} )^{-4/3}  \\  \\
    = {\rm T_\sun^{2/3}  } e \ 
(P_b / 2 \pi)^{1/3} \ \frac {  m_{2;  \langle\dot{\omega}\rangle,\gamma}  }    {\rm M_\sun} \times\\ 
 \   \left( \frac { m_{1;  \langle\dot{\omega}\rangle,\gamma} + 2 m_{2;  \langle\dot{\omega}\rangle,\gamma}   }  
  {\rm M_\sun}  \right)
 \   \left( \frac {(m_{1;  \langle\dot{\omega}\rangle,\gamma} + m_{2;  \langle\dot{\omega}\rangle,\gamma}      }  
  {\rm M_\sun}  \right)^{-4/3}.
\end{multlined}
\end{equation}
Solving simultaneously for  the two component masses, we find that   
$m_{1;  \langle\dot{\omega}\rangle,\gamma}=  
1.438  \pm  0.001\  {\rm M}_{\sun}$ and
$m_{{2;  \langle\dot{\omega}\rangle,\gamma}}=  	1.390  \pm 0.001\   {\rm M}_{\sun}$.  
These values agree with 
WNT to within $2\sigma$, while our precision is 
poorer due to a less-precisely determined $\gamma$ (see \S\ref{sec:results}).
Furthermore, Newtonian physics then yields
an additional quantity from the derived masses and our $x$ and  $P_{\rm b}$  
measurements:
$(\sin i)_{ \langle\dot{\omega}\rangle,\gamma}=	0.7327\pm 0.0004$ (or,
equivalently, $|\cos i|_{ \langle\dot{\omega}\rangle,\gamma}=	0.6806\pm 0.0004$).
These values are presently much more precise than the
$m_{2}$ and $\sin i$ or $|\cos i|$
 values determined directly from the 
Shapiro propagation delay  measurements of \S\ref{sec:shapiromeas}.

\subsection{Toward the First  Published Measurement of  the 
Relativistic Orbital  Shape 
Correction,  $\delta_\theta$, in any System}
\label{sec:shapeobs}

We have successfully measured the apparent
post-Keplerian orbital shape correction term,  $\delta_{\theta}^{\mbox{obs}}$ (see Table  
\ref{table:orbfits}).  As noted in \S\ref{sec:deltathetatheory}, this observed value 
must be corrected for a comparable aberration
signal  $\epsilon_{\rm A}$ (see Eq.  \ref{eqn:deltathetaobs}).  Geodetic 
spin-precession modeling  of this system 
should in principle determine the necessary  aberration parameters
by specifying the spin axis orientation (specifically, its polar angles $\eta$ and $\lambda$) 
over time (see Eq. \ref{eqn:A}).
 However, we find that the currently available pulseshape variation fits
 \citep{k98,wt02,cw08}  yield inconsistent  solutions for these parameters.  
Consequently, although
we have successfully measured $\delta_{\theta}^{\mbox{obs}}$,
it is not yet possible to determine $\delta_\theta^{\rm intr}$ nor hence to use its  
measured value as an 
additional test of relativistic gravitation.  

Despite our current inability to {\it{measure}} $\delta_\theta^{\rm intr}$, we can 
determine its {\it{expected}} value $\delta_\theta^{\rm GR}$ via a general relativistic calculation
(DD; DT92):
 \begin{align}
\delta_{\theta}^{\rm GR} 
= &\frac{\rm G^{2/3}}{c^2 } \left(\frac{P_{\rm b}  }{2 \pi} \right)^{-2/3} 
(m_{1} + m_{2})^{-4/3}
\ \ \left\{  \frac{7}{2}    m_{1}^2  +6 m_{1} m_{2} + 2 m_{2}^2    
\right\} \nonumber \\
= &{\rm T_\sun}^{2/3}  \left(\frac{P_{\rm b}  }{2 \pi} \right)^{-2/3} 
\left( \frac{m_{1} + m_{2}} 
{\rm M_\sun} \right)^{-4/3} \ \ \ \ \ \times \nonumber  
\end{align}
\begin{equation}
 \ \ \ \ \ \ \ \ \  \ \ \ \ \ \ \ \ \ \ \ \ \ \ \ \ \ \ \ \ \ \ \ \ \ \ \ \ \ \ \ \left\{  
\frac{7}{2}   \left( \frac{m_{1} }{\rm M_\sun} \right)^2  
+6  \left(\frac{m_{1} m_{2} }{\rm M_\sun}\right) 
+2 \left( \frac{m_{2}} {\rm M_\sun} \right)^2  
\right\},
\end{equation}
yielding $\delta_\theta^{\rm GR}=  (6.187\pm 0.001) \times10^{-6}$. 
 Eq.  \ref{eqn:deltathetaobs} can then be inverted to give 
 the aberration parameter $\epsilon_{\rm A}= (2.2\pm 2.5) \times10^{-6}$. 
This timing-derived value of $\epsilon_{\rm A}$  will provide a modest
consistency 
check on future geodetic precession modeling.

\subsection{Implications of Fits for  the First Time-Derivatives of $e$ and $x$  }
\label{sec:xdotedotobs}

As noted above, Eqs. \ref{eqn:edot} and \ref{eqn:xdot} demonstrate that the 
successful measurement 
of $\dot{e}^{\rm obs}$ and $\dot{x}^{\rm obs}$ would  lead to constraints on other 
quantities of interest.  
It is therefore useful to further investigate the various terms composing these equations.  

The first (aberration) term of both equations, $d \epsilon_A / dt$, was defined in Eqs.  
 \ref{eqn:epsilonA} and  \ref{eqn:depsilonAdt}.  However, as discussed in 
 \S\ref{sec:shapeobs} for $\epsilon_A$, additional progress in understanding the pulsar's   
 spin axis orientation  is needed before $d \epsilon_A / dt$ can be  confidently determined. 
Nevertheless, our current geodetic precession modeling suggests that its value is in the 
$\sim10^{-15}\  \rm s^{-1}$ range, and varying with  spin-precessional phase.

\subsubsection{Constraints from $\dot{e}^{\rm obs}$}
\label{sec:edot}
The   second and final term of  Eq.  \ref{eqn:edot} 
 involves the time evolution of $e$ induced by gravitational wave (``GW'') emission. 
For the Eq.  \ref{eqn:edot} second term, \citet{p64} shows that this term, 
\begin{equation}
\begin{multlined}
\left(\frac{ \dot{e} }{e}\right)^{\mbox{GW}} 
= - \frac{304}{15} \frac{{\rm G}^3}{c^5 a_{\rm R}^4} \ \mu \  (m_{1} + m_{2})^2  \times \\ 
        (1-e^2)^{-5/2}   \ \left(1+\frac{121}{304}e^2\right)  \\
 = - \frac{304}{15} \  {\rm T_\sun}^{5/3} 
  \frac { \frac{m_{1} } {\rm M_\sun} \frac{  m_{2}  }  {\rm M_\sun}  }
 { (\frac{m_{1} + m_{2} } {\rm M_\sun}   )^{1/3}   }
  \left(\frac{P_{\rm b}  }{2 \pi} \right)^{-8/3}   \times \\
    \ \              (1-e^2)^{-5/2}   \ \left(1+\frac{121}{304}e^2\right)  \\ \\
 = -2.9 \times 10^{-17}\  \rm s^{-1},
 \end{multlined}
\end{equation}
with $\mu$ the reduced mass. This term is negligible compared
 to the expected value of $d \epsilon_A / dt$, except at fortuitous precessional phases 
 where the latter can drop to zero.  Consequently, the
 $d \epsilon_A / dt$ term dominates Eq. \ref{eqn:edot}, so that a successful measurement of 
 $(\dot{e} / e)^{\rm obs}$ would provide a unique measurement of $d \epsilon_A / dt$.  
 Unfortunately, our fitted value of $(\dot{e} / e)^{\rm obs}$ is currently not significantly 
 different from zero, although  its upper limit is in
 the $10^{-15}\  \rm s^{-1}$ range (see Table \ref{table:orbfits}) expected for $d \epsilon_A / dt$.

\subsubsection{Constraints from $\dot{x}^{\rm obs}$}

The  second term  in Eq. \ref{eqn:xdot} delineates the gravitational wave-induced 
orbital shrinkage rate,
which can be evaluated from measurables via
\begin{equation}
\left(\frac{ \dot{a}_{1} }{{a}_{1}}\right)^{\mbox{GW}} = 
\frac{2}{3} \left(          \frac{ \dot{P}_{\rm b} }{  {P}_{\rm b} }\right)^{\mbox{GW}}=
 -5.7 \times 10^{-17} \ {\rm s}^{-1}.
\end{equation}

The third (spin-orbit) term of Eq. \ref{eqn:xdot} varies approximately sinusoidally on the geodetic
precession timescale, with an amplitude of $\sim \pm 3 \times 10^{-15} \ {\rm s}^{-1}$.  Details
await a robust determination of geodetic precession parameters (see Eq. \ref{eqn:didt}). 

The fourth \citep{k96} term of Eq.  \ref{eqn:xdot} has a maximum amplitude\footnote{The exact 
value depends on the unknown alignment on the sky of the line of nodes.} 
of $\sim 2.3 \times 10^{-16} 
\ {\rm s}^{-1}$;
while the fifth and final term, 
\begin{equation}
-\frac{\dot{D}}{D} = +\frac {   \dot{P}_{\rm b,\ gal}  }   {  P_{\rm b}        }    
= -1 \times 10^{-18}  \ {\rm s}^{-1}
\label{eqn:DdotoverD}
\end{equation}
(see \S\ref{sec:obsintPbdot} for details on the calculation of $\dot{P}_{\rm b,\ gal}$). 

In summary, the first (aberration) and third (spin-orbit) terms dominate Eq. \ref{eqn:xdot}, so  
all others may be ignored. However, neither of these two terms is currently 
accurately determinable. 
We do have a marginally significant measurement of $\dot{x}^{\rm obs}$ (see
Table \ref{table:orbfits}).  Consequently,  if either of the two terms becomes well-determined 
in the future along with an improved value of $\dot{x}^{\rm obs}$, then the other 
term will also become 
accessible.  For example, there are two possible paths toward determination of 
the aberrational term:
First, additional  geodetic precession observations and modeling should better constrain
 $d \epsilon_A / dt$;
and second, additional observations could better determine $\dot{e}^{\rm obs}$, which, 
as noted 
in \S\ref{sec:edot},
would then be equivalent to a measurement of $d \epsilon_A / dt$.  At this point, the spin-orbit 
term would
be calculable,  leading to an exciting measurement of the pulsar's moment of inertia, 
which has
important implications for neutron star equations of state \citep{ls05}. With the measurement 
precision of $\dot{x}$ and  $\dot{e}$
 improving with time {\it t} as $t^{-3/2}$, another decade or so of observations are required.
Unfortunately, geodetic spin axis precession may cause the pulsar to disappear before that time.

\section{Tests of Relativistic Gravitation}
\label{sec:tests}

The determination  of seven
particular independent quantities suffices to fully determine the dynamics of 
a binary system within the context of a particular theory of gravitation.  For example, the 
most accurate
determination of component masses and orbital inclination in \S\ref{sec:bestmasses}
 and of $\delta_{\theta}^{\rm GR}$ in 
\S\ref{sec:shapeobs} depend upon subsets of 
the first seven measurements listed in Table \ref{table:orbfits}.

Consequently, any additional measurement constitutes an independent test of relativistic
gravitation under strong-field conditions.  In the following two sections, we delineate 
relativistic gravitational 
tests via measurements of gravitational radiation emission and of the Shapiro 
gravitational propagation delay, respectively.

\subsection{Gravitational Radiation Emission and $\dot{P}_{\rm b} $ }
\label{sec:gravrad}
	
Gravitational radiation emission should cause the orbit to decay as orbital 
energy is radiated away.
The quantity $\dot{P}_{\rm b}^{\rm GR}$ is the resulting orbital period 
derivative expected from a
general relativistic calculation of this phenomenon  \citep{pm63}:
\begin{equation}
\begin{multlined}
\dot{P}_b^{\rm GR} =  -\frac{192\,\pi\,G^{5/3} } {5\,c^5}
\left(\frac{P_b}{2\pi}\right)^{-5/3} 
  \left(1 + \frac{73}{24} e^2 + \frac{37}{96} e^4\right) (1-e^2)^{-7/2}   \\ \\
\times\ m_{1}\,m_{2} \ \ (m_{1}+m_{2})^{-1/3} \\ \\
 = - \frac{192\,\pi } {5} \ T_\sun^{5/3}\ \left(\frac{P_b}{2\pi}\right)^{-5/3}  
   \left(1 + \frac{73}{24} e^2 + \frac{37}{96} e^4\right) (1-e^2)^{-7/2}  \\
\times\  \left(  \frac{ m_{1} } {\rm M_\sun} \right) \left(  \frac{m_{2}  }{ {\rm 
M_\sun} }\right)    \left(\frac{ m_{1}+m_{2} } {\rm M_\sun} \right)^{-1/3} \\
\end{multlined}
\label{eq:pbdottheo}
\end{equation}

Inserting our measured and derived values and their uncertainties into Eq. \ref{eq:pbdottheo}
\footnote{This value may also be calculated directly from  
the first seven 
orbital measurables in Table   \ref{table:orbfits} alone, without
the use of derived quantities such as the masses (DT91).},
we find that $\dot{P}_b^{\rm GR} =(-2.40263 \pm 0.00005)  \times 10^{-12}$. 
To verify our  estimate of the error in $\dot{P}_b^{\rm GR}$  that was derived
via propagation of uncertainty, we also employed
a Monte-Carlo method with Cholesky decomposition of the covariance matrix. In this fashion, we
 simulated the joint normal distribution
of measured parameters \{$\gamma,\dot{\omega},P_{\rm b}, e$\}, and	
then constructed a histogram of $1000000$ derived 
 $\dot{P}_b^{\rm GR}$ and inferred the uncertainty therefrom.

Consequently, we find that 
 \begin{equation}
  \frac{  \dot{P}_b^{\rm intr}  }{  \dot{P}_b^{\rm GR}  }=
  \frac{ (-2.398 \pm 0.004)  \times 10^{-12}  }
          { (- 2.40263 \pm 0.00005)  \times 10^{-12}  } = 	0.9983 \pm 0.0016.
 \end{equation}

\begin{figure}
\vspace{0.6in}
\includegraphics[scale=0.45,trim={0in 0.4in 1.8in 1.5in}  ]
{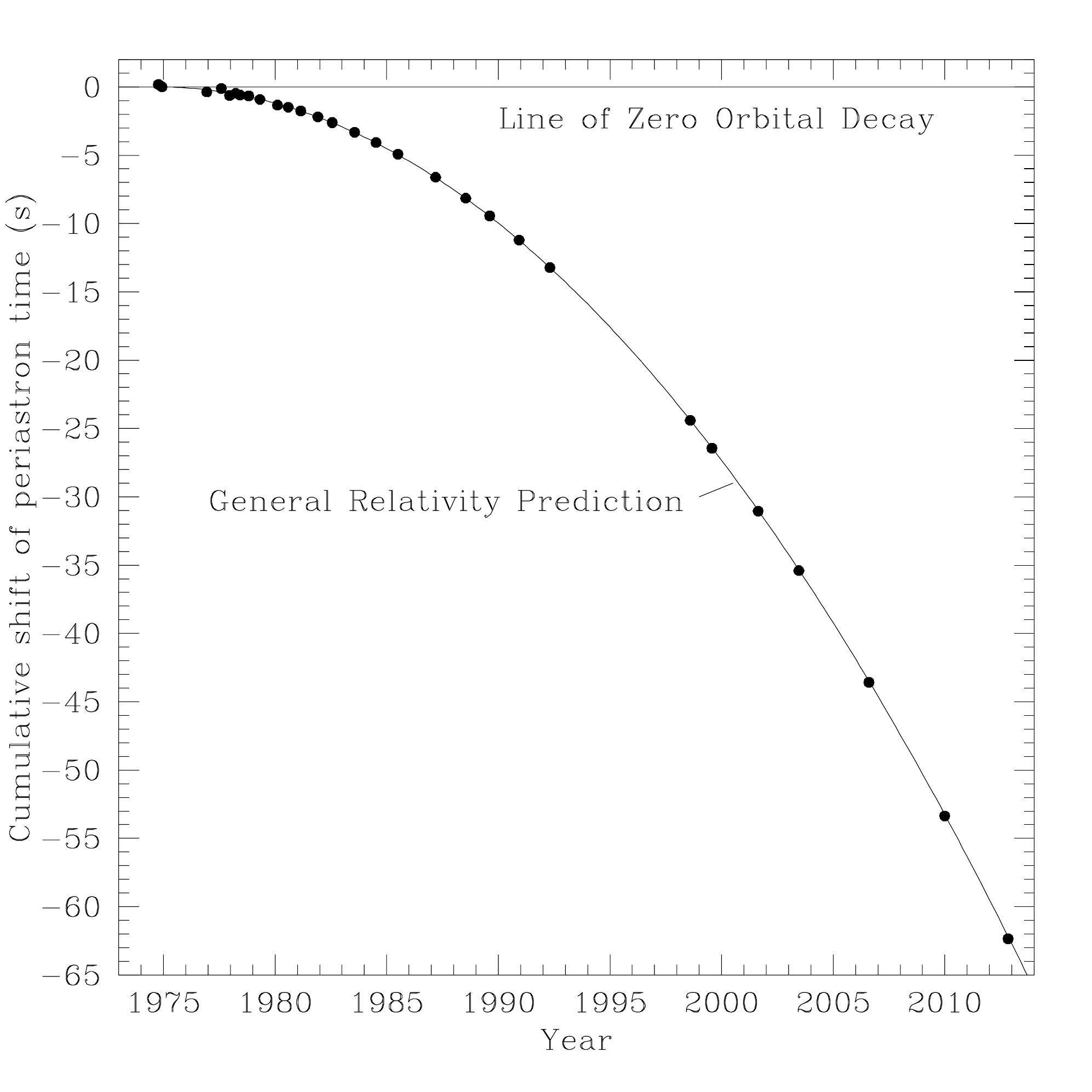} 
\caption{The orbital decay of PSR B1913+16 as a function of time. The curve represents the
 orbital phase shift  expected from gravitational wave
emission according to General Relativity. The points, with error bars too small to show, 
represent our measurements.}
\label{fig:parab}
\end{figure}

This result demonstrates that the system is losing energy to gravitational radiation within 
$\sim1\sigma$  of the rate predicted by general relativity 
 (see also Fig.  \ref{fig:parab} and the red curve in Fig. \ref{fig:massmass}).  The above
 number represents a significant improvement over the value determined by WNT, $0.997 \pm 0.002$, which 
 represented a 1.8$\sigma$ discrepancy between our measurements and general relativity. 
 Interestingly, the new galactic parameters of \citet{ret14} are the principal reason for the 
 improvement (via a change in $\dot{P}_{\rm b}^{\rm gal}$), while our measured values 
 themselves changed little.
    
  In addition to  confirming general relativistic radiation damping at this level, our result 
  rules out large parameter spaces in
  plausible  scalar-tensor  theories of gravity.  In recent years, however, other  pulsars in neutron 
  star-white dwarf binary systems have
  overtaken PSR B1913+16 in constraining these alternatives \citep{fet12}.
  
  DD92 point out that this test is a ``mixed'' strong-field probe in that it involves 
  a combination of  
  radiative effects (via $\dot{P}_{b}^{\rm obs}$) and quasi-static phenomena (through 
  $\langle\dot{\omega}\rangle$ and
  $\gamma$,   whose values are needed  in order to make a prediction of the expected 
  $\dot{P}_{b}^{\rm GR}$).  Consequently, additional tests, such as those described in the
  next section, that probe different aspects of strong-field gravitation, 
  are also useful in constraining viable alternatives to general relativity.  
  

 \subsection{Shapiro Gravitational Propagation Delay  }
\label{sec:testshap}
Each of the two newly measured Shapiro parameters represents another 
independent test of relativistic gravitation. As with the $\dot{P}_{b}$ test of 
\S\ref{sec:gravrad}, the 
Shapiro tests also require the complementary measurement of  
$\langle\dot{\omega}\rangle$ and $\gamma$ in order to make a testable
prediction for the value of the Shapiro parameters.  In this case, unlike the $\dot{P}_{b}$ test,
all of the post-Keplerian
quantities probe quasi-static phenomena in strong fields. While  
Shapiro parameters have 
already been measured in several other binary systems, it is especially useful to constrain 
theories via systems such as this one  and PSRs B1534+12 and J0737-3039A,
where at least three ``excess'' post-Keplerian parameters 
beyond $\langle\dot{\omega}\rangle$ 
and $\gamma$  (one gravitational radiation parameter and  two Shapiro quantities) are 
measurable.  Although the precision of  
 the binary pulsar Shapiro parameter measurements is well below their measurement precision 
 in the weak solar 
 gravitational field, it is this simultaneous determination of several parameters in strong-field 
 conditions in each of these binary pulsar systems that leads to the important constraints on 
 relativistic theories of gravitation.
	
In the DD formulation of the Shapiro delay within general relativity,  the Shapiro
measurables $s$ and $r$ map  directly onto  $\sin i$ and $m_2$, respectively. 
 Hence we can test General Relativity by comparing the Shapiro determination 
 of $\sin i\ (\equiv s)$
 with that determined from $\langle \dot \omega \rangle$ and 
 $\gamma$ (called $\sin i_{\langle \dot \omega \rangle, \gamma}$;  see \S\ref{sec:bestmasses}):
\begin{equation}	
\frac{s} { \sin i_{\langle \dot \omega \rangle, \gamma}   }= 
\frac{ 0.68\substack{+0.10\\-0.06}}{0.7327\pm 0.0004}=\   
0.93\substack{+0.14\\-0.07} . \\
\end{equation}
Similarly, we can test General relativity by comparing 
the Shapiro determination of $m_2\ (\equiv r\  {\rm M}_\sun/ {\rm T}_\sun$)  with that 
determined from $\langle \dot \omega \rangle$ and $\gamma$:

\begin{equation}	
\frac{r \  {\rm M}_\sun/ {\rm T}_\sun } { m_{2; \langle \dot \omega \rangle, \gamma}  }= 
 \frac{(1.95 \substack{+0.55\\-0.71})\  {\rm M}_\sun}{(1.390 \pm 0.001)\rm M_\sun}=\ 
1.40 \substack{+0.40\\-0.51} .
\end{equation}

\begin{figure}
\vspace{0.3in}
\includegraphics[scale=0.5,angle= 0, trim={0.5in 1.4in 0in 3in}  ]
{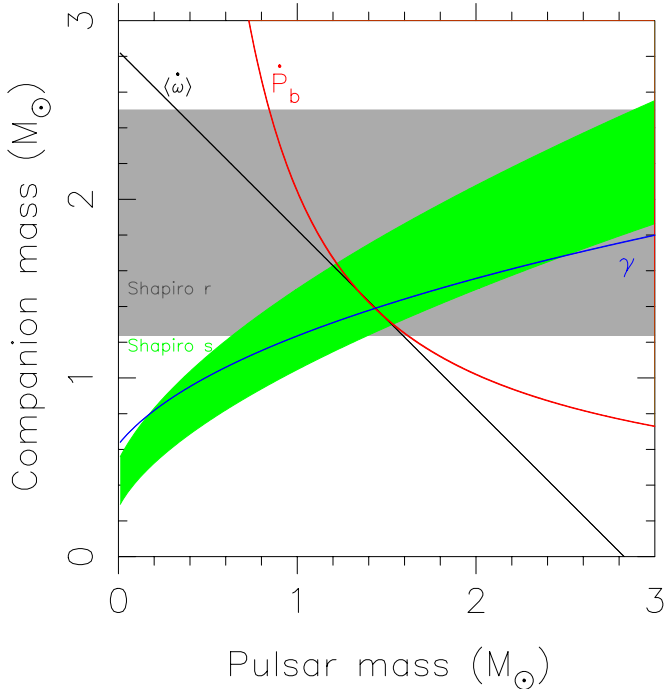} 
\caption{ Constraints on the masses of the pulsar and the companion as a function of five 
post-Keplerian measurables, within the 
context of General Relativity. The width of each curve represents $\pm 1\sigma$ error bounds. 
The mutual near-intersection of all curves  illustrates the agreement of our
observations with general relativity in the strong-field conditions at the binary system.
}
\label{fig:massmass}
\end{figure}

The consistency  (albeit with a rather low level of precision)
 of  these  Shapiro determinations of $\sin i$ and $m_2$ with 
those measured via 
the other post-Keplerian terms, and hence their confirmation of general relativity, 
is also  graphically depicted in Fig. \ref{fig:massmass}.	The Shapiro terms 
have also been
measured in several other binary pulsar systems with higher precision, and
have also been shown to be in agreement
with general relativity.

\section{Conclusions}
\label{sec:concl}
We report here on the measurements and relativistic analyses of 
9309 TOAs in over thirty years of high-quality Arecibo
data on binary pulsar PSR B1913+16. 	
We fitted for a number of previously 
unmeasurable parameters for the first time in this system (and in
one case, for the first time anywhere), which enabled us to significantly
advance our relativistic analyses of the system. 
We provide our newest measurents or derivations 
of all relevant physical quantities of the binary system, with the exception of $\Omega$, 
the position angle of the line of nodes.  
We rigorously ascertained the uncertainties in the fitted and derived parameters. 
Having fully characterized the system, we 
proceeded to use it in several tests of general relativity in strong-field conditions.

We have measured a gravitational radiation-induced orbital period decrease 
whose rate agrees with the general relativistic expectation to	
within $\sim1\sigma$, which is closer than found by WNT, largely as a result of
an improved galactic correction resulting from more accurate galactic parameters \citep{ret14}.   

\begin{deluxetable}{lll}
\tablecolumns{3}
\tablecaption{Comparison of Gravitational Radiation-Induced Orbital Decay with GR Prediction in Binary Pulsars}
\tablehead{
\colhead{PSR}  &   \colhead{$\dot{P}_{\rm b}^{\rm intr} / \dot{P}_{\rm b}^{\rm GR}$} &  \colhead{Ref.} \\
}
\startdata

J0348+0432 & $1.05 \pm 0.18$ & \citet{aet13} \\
J0737-3039 &$1.003 \pm 0.014$ & \citet{ket06} \\
J1141-6545& $1.04 \pm 0.06$&  \citet{bbv08}\\
B1534+12  &$0.91 \pm 0.06$&  \citet{set02} \\
J1738+0333   &$0.94 \pm 0.13$&  \citet{fet12} \\
J1756-2251   &$1.08 \pm 0.03$&  \citet{fet14} \\
J1906+0746   &$1.01 \pm 0.05$\tablenotemark{a}&   \citet{vet15} \\
B1913+16      &$0.9983 \pm 0.0016$&   This work \\
B2127+11C    &$ 1.00\pm 0.03$&  \citet{jet06} \\
\enddata 
\tablenotetext{a}{Assumes negligible proper motion.}
\label{table:Pbdots}
\end{deluxetable}

 Similar orbital decay tests have now been performed with  several other binary pulsars 
 (see Table \ref{table:Pbdots} for published
 measurements of  $\dot{P}_{\rm b}^{\rm intr} / \dot{P}_{\rm b}^{\rm GR}$ ).   
 The  orbital  decays of PSRs  J0348+0432,
 J0737-3039, J1141-4565,  J1738+0333, J1906+0746, B1913+16, and   
 B2127+11C  all exhibit agreement between observation and general relativity to within (or very
 close to) the authors' stated uncertainties.  PSR B1913+16 currently has the most
 precise such determination, and interferometric parallax measurements, 
 currently in progress, will hopefully further tighten the precision. 
 
Among the other two, PSR B1534+12 and PSR J1756-2251,
  various systematic effects such as an incorrect distance in the galactic 
  acceleration correction may 
  explain the small  observed discrepancies, although it is  possible that an
incompleteness of general relativity or some unknown physical effect is responsible.  
See  the work of \citet{fet14}  for an especially thorough description of the most
significant deviation of the orbital decay rate from the general relativistic prediction, found in
PSR J1756-2251.

 Our  new (for this system) measurements of Shapiro gravitational propagation delay 
 parameters represent  two additional tests of relativistic gravitation, and are fully 
consistent with general relativity, although their relative precision is currently far lower than the 
orbital decay test. This binary now joins several other systems, 
including PSRs J0737-3039A, 
B1534+12, and J1756-2251, in each
providing at least  three independent tests of relativistic, strong-field  gravitation.

We have also marginally
measured the orbital shape parameter $\delta_{\theta}$ for the first time anywhere, 
but its intrinsic value is corrupted by a comparable, undetermined aberration delay. 
Future geodetic spin-orbit precession measurements should lead to an accurate 
characterization of the aberration and then an additional
relativistic gravitational test via the comparison of the aberration-corrected 
$\delta_{\theta}^{\rm intr}$ with $\delta_{\theta}^{\rm GR}$.

In addition, we  fitted for the time derivative of orbital eccentricity $e$ and 
projected semimajor axis of the pulsar orbit, $x$, and we achieved 
an upper limit on the former and
a marginal detection of the latter.  We discussed and
quantified the various physical phenomena that can contribute to these parameters.
Unless the pulsar disappears in the next few years due to geodetic spin axis precession,
future timing observations
should better define these quantities, allowing for a determination of the pulsar's 
moment of inertia $I_1$.

We have  placed  online$^2$ 
a subroutine and modifications to  the TEMPO TOA fitting 
software, which codes the \citet{fw10} parametrization of the Shapiro delay for
high-eccentricity  binary pulsars such as the PSR B1913+16 system.
An  online companion of this paper provides the TEMPO input files and TOAs 
upon which these analyses are based.  The same data are on arXiv, and at zenodo:  \dataset[10.5281/zenodo.54764]{http://dx.doi.org/10.5281/zenodo.54764}.

\acknowledgements{Much of this experiment was pioneered by J. H. Taylor, to
whom we owe our deepest thanks. D. J. Nice assisted with observing and analyses,
and A. A. Chael assisted in developing the FW analysis package.
The authors gratefully acknowledge 
financial support from the  US National Science Foundation.  The Arecibo
Observatory is operated by SRI International under a cooperative agreement 
with the National Science Foundation (AST-1100968), and in alliance with 
Ana G. Mendez-Universidad Metropolitana, and the Universities Space 
Research Association. }

{\it Facility:} \facility{Arecibo}

\end{document}